\documentclass[aps,prl,twocolumn,superscriptaddress]{revtex4-2}
\usepackage{graphicx}
\usepackage{dcolumn}
\usepackage{bm}
\usepackage[latin1]{inputenc}
\usepackage{textcomp}
\usepackage{amsmath}
\usepackage{color}

\begin{document}

\title{Hydration at highly crowded interfaces}

\author{Christopher Penschke}
\email[]{penschke@uni-potsdam.de}
\affiliation{Department of Chemistry, University of Potsdam, Karl-Liebknecht-Str.\ 24-25, D-14476 Potsdam-Golm, Germany}

\author{John Thomas}
\affiliation{Faculty of Physics and Center for Nanointegration (CENIDE), University of Duisburg-Essen, D-47048 Duisburg, Germany}

\author{Cord Bertram}
\affiliation{Faculty of Physics and Center for Nanointegration (CENIDE), University of Duisburg-Essen, D-47048 Duisburg, Germany}
\affiliation{Department of Chemistry and Biochemistry, Ruhr-Universit\"{a}t Bochum, Universit\"{a}tsstr.\,150, D-44801 Bochum, Germany}

\author{Angelos Michaelides}
\affiliation{Yusuf Hamied Department of Chemistry, University of Cambridge, Lensfield Road, Cambridge CB2 1EW, United Kingdom}

\author{Karina Morgenstern}
\affiliation{Department of Chemistry and Biochemistry, Ruhr-Universit\"{a}t Bochum, Universit\"{a}tsstr.\,150, D-44801 Bochum, Germany}

\author{Peter Saalfrank}
\affiliation{Department of Chemistry, University of Potsdam, Karl-Liebknecht-Str.\ 24-25, D-14476 Potsdam-Golm, Germany}
\affiliation{Department of Physics and Astronomy, University of Potsdam, Karl-Liebknecht-Str.~24-25, D-14476 Potsdam-Golm, Germany}

\author{Uwe Bovensiepen}
\email[]{uwe.bovensiepen@uni-due.de}
\affiliation{Faculty of Physics and Center for Nanointegration (CENIDE), University of Duisburg-Essen, D-47048 Duisburg, Germany}

\date{\today}

\begin{abstract}
Understanding the molecular and electronic structure of solvated ions at surfaces requires an analysis of the interactions between the surface, the ions, and the solvent environment on equal footing. Here, we tackle this challenge by exploring the initial stages of Cs$^+$ hydration on a Cu(111) surface by combining experiment and theory. Remarkably, we observe ``inside out'' solvation of Cs ions, i.e, their preferential location at the perimeter of the water clusters on the metal surface. In addition, water-Cs complexes containing multiple Cs$^+$ ions are observed to form at these surfaces. Established models based on maximum ion-water coordination and conventional solvation models cannot account for this situation and the complex interplay of microscopic interactions is the key to a fundamental understanding.
\end{abstract}

\maketitle

The interplay of screening, local charge accumulation and high electronic density of states at metal surfaces is decisive for understanding the fundamental aspects of surface reconstruction and reactions and, linked to both, heterogeneous catalysis \cite{noerskov_2014, thomas_thomas_2015}.
This is particularly pronounced for, but not limited to, metal-electrolyte interfaces, because screening and local reorganization occurs in this situation not only in the metal, but in the electrolyte as well \cite{gonella_review_2021}. Prominent examples of such interfaces include alkali ions in aqueous solution approaching metal surfaces \cite{schmickler_approach_2022,xi_distribution_2020}. Even without an extensive liquid phase, solvation of adsorbates may occur due to the presence of water, either as a reactant or as part of the environment. The different types of interactions (ion--surface, ion--solvent, solvent--surface, solvent--solvent, ion--ion) contribute to a wide range of operational parameters, leading to a multifaceted landscape of potential applications. However, this complexity also impedes understanding of such interfaces.

Surface science studies at defined model systems provide a well-established approach to analyze the fundamental, microscopic interactions since they promise insights regarding competing or cooperative effects in general \cite{kolb_electrochemical_2001}. One such model system is Cs$^+$ on Cu(111), which has been investigated for a range of catalytic conversions \cite{campbell_model_1987,shimizu_supramolecular_2014,resasco_promoter_2017,monteiro_absence_2021,hamlyn_structure_2020}, including reactions involving water (e.g., the water-gas shift reaction).

The interaction between adsorbate and metal surface affects both, the geometric and the electronic structure at the interface \cite{henderson_interaction_2002}. It is well-known that the  most stable clusters in bulk water consist of single alkali ions surrounded by four (Li$^+$) to eight (Rb$^+$, Cs$^+$) water molecules \cite{mahler_study_2012}. Close to a transition metal surface, the solvation structures may be entirely different \cite{weber_structural_2019,lucht_imaging_2018}. Model systems with reduced complexity (i.e., adsorption at submonolayer coverage under ultra-high vacuum conditions) provide valuable contributions to understanding the interface and the fundamental interactions determining its structure. In addition, such model studies can also reveal changes in the electronic structure of the surface \cite{dahl_electronic_2001}. The formation of alkali ions upon adsorption of neutral alkali atoms is due to an electron transfer to the surface accompanied by the formation of unoccupied, short-lived resonances, which are decisive for the occurrence or non-occurrence of photoreactions \cite{petek_surface_2002}. For example, photoexcitation into the Cs $6s$ resonance on Cu(111) leads to an increase of the Cs-Cu bond distance  \cite{ogawa_phase_1999}, but photodesorption is unlikely because of the low cross section under experimental conditions \cite{petek_real-time_2000,kroner_laser-induced_2007}. While changes in energy and lifetime of the alkali resonances on Cu(111) by solvation were experimentally detected \cite{meyer_ultrafast_2015,thomas_competition_2021}, a microscopic understanding of the complex interactions of water and alkali ions on metal surfaces remains elusive.

In this work, we investigate the relationship between the geometric and electronic structure of Cs$^+$ ions coadsorbed with water on a Cu(111) surface. Combining density functional theory (DFT), scanning tunneling microscopy (STM), and  two-photon photoelectron emission (2PPE), we show how the relative strengths of water-ion and water-water interactions lead to coverage-dependent changes in the structure of the co-adsorbates. We find that clusters crowded with ions form, with the ions solvated at the perimeter of the water clusters. The results also suggest that the local ion/water concentration causes pronounced energetic shifts and splitting of the Cs resonance, which may significantly affect the properties of the interface.

\begin{figure}[t]
\includegraphics[width=0.99\columnwidth]{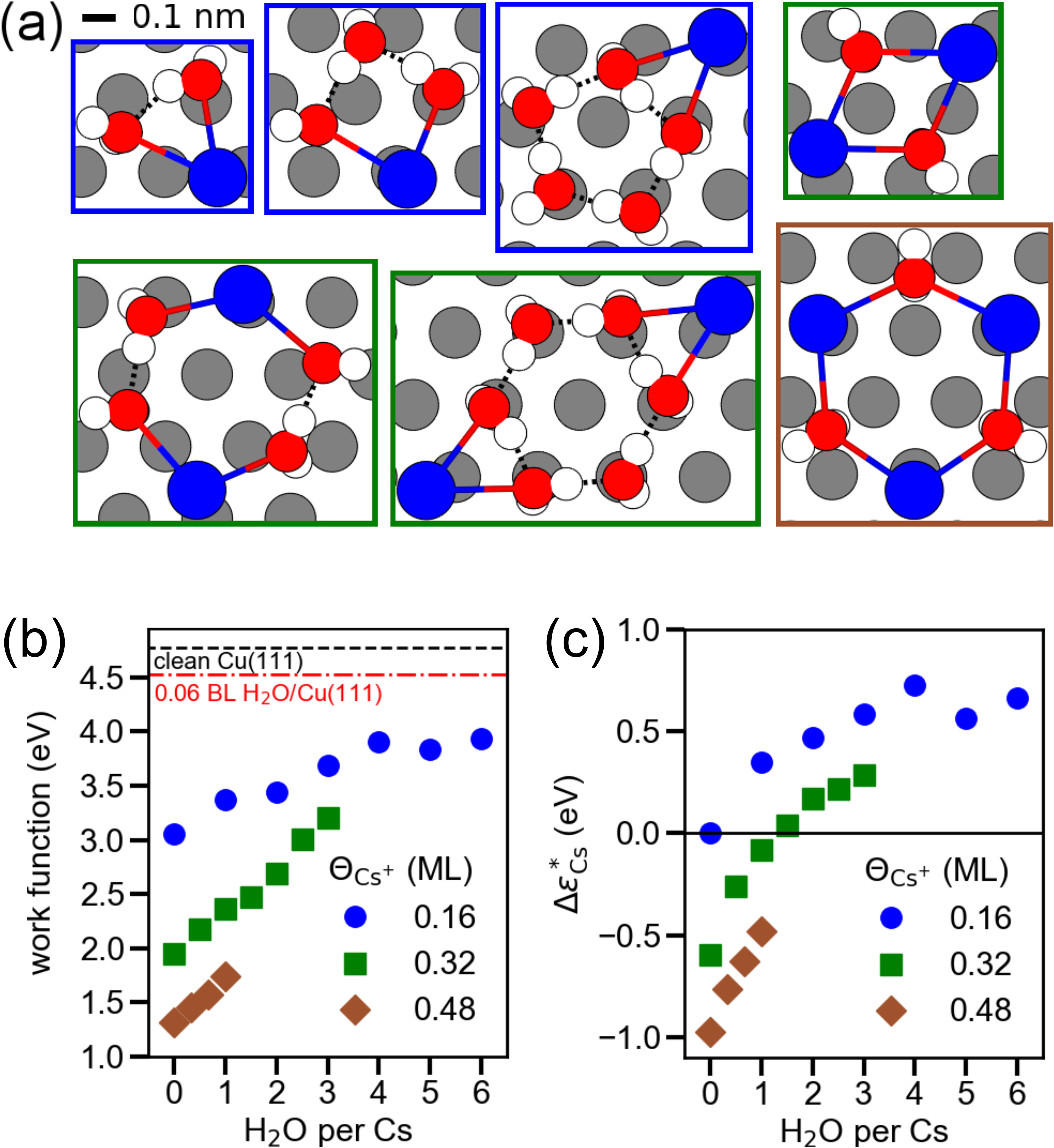}
\caption{ (a) Clusters of one Cs$^{+}$ ion with two, three, and five water molecules, two Cs$^{+}$ ions with two, four and six water molecules, and three Cs$^{+}$ ions with three water molecules obtained by DFT calculations. Cs, Cu, O, and H are depicted in blue, grey, red, and white, respectively. A complete overview of the structures is available in the Supplemental Material (Figs.\ S5-S7). (b) Computed work function as a function of the number of water molecules per Cs$^{+}$, for adsorbed clusters containing one (blue circles), two (green squares), and three (brown diamonds) Cs$^{+}$ ions, which correspond to coverages of 0.16, 0.32 and 0.48~ML, respectively. The work function of clean Cu(111) and 0.06~BL H$_2$O/Cu(111) (one adsorbed water molecule per 5$\times$5 unit cell) is indicated by black and red lines, respectively. (c) Computed band center shift of the unoccupied Cs $6s$ states, $\Delta \epsilon_{\mathrm Cs}^*$, relative to a single adsorbed Cs$^{+}$ ion (0.16 ML); see Supplemental Material.
\label{image_dft}}
\end{figure}

To calculate the Cs$^{+}$-water clusters on Cu(111) for different Cs$^+$ and H$_2$O coverage we used the exchange-correlation functional by Perdew, Burke, and Ernzerhof (PBE) \cite{perdew_generalized_1996} together with the D3 dispersion correction \cite{grimme_consistent_2010,grimme_effect_2011} as implemented in VASP \cite{kresse_efficiency_1996,kresse_efficient_1996}. The Cu(111) surface was modelled by 5$\times$5 surface unit cells, in which the different clusters are placed individually. Thus, the clusters are defined by the number of Cs$^+$ ions (1 to 3) and water molecules (0 to 6) per unit cell. One Cs$^+$ per unit cell corresponds to a coverage of 0.16~monolayers (ML), defined with respect to a closed-packed (2$\times$2) monolayer. The water coverage is given in fractions of a bilayer (BL), which is a closed hexagonal layer with 2 water molecules per 3 surface atoms \cite{henderson_interaction_2002}. See the Supplemental Material for more details.

The Cs$^{+}$-water and the water-water interactions are of comparable strength. For instance, the agglomeration energy of a Cs$^{+}$ ion and a water molecule on Cu(111) is -0.36~eV, compared to a value of -0.28~eV for two water molecules. This competition between ion-water and water-water interactions has been discussed in the context of gas-phase Cs$^{+}$-water clusters \cite{kolaski_structures_2007}. In contrast to the typically three-dimensional gas-phase clusters with many Cs-water bonds, flat structures are more favourable on Cu(111) due to the strong adsorbate-surface bonds. Because of the large size of Cs$^{+}$, it is difficult to build flat clusters with both, many Cs$^{+}$-water and many hydrogen bonds. Remarkably, we find that the most stable structures are hydrogen-bonded water clusters with Cs$^{+}$ located at the perimeter or as part of a ring, see Fig.\ \ref{image_dft}(a). We emphasize that structures with Cs$^{+}$ in the center of surrounding water with maximal coordination on Cu(111), which were analyzed in \cite{perez_paz_hydrated_2021}, are found to be less stable, see Supplemental Material.

Increasing the amount of water per Cs changes the electronic structure. Up to four water per Cs$^{+}$ ion, the work function increases by 1~eV, see Fig.\ \ref{image_dft}(b), and the band center of the unoccupied Cs$^{+}$ 6s states shifts to higher energies by 0.8~eV, see Fig.\ \ref{image_dft}(c) and Supplemental Material, Fig.\ S4. We also investigated clusters with multiple Cs$^{+}$ ions. The cluster structures are similar to those with one Cs$^{+}$ ion, i.e., Cs$^{+}$ is located at the perimeter of water clusters or part of a ring. The Cs-Cs distances are 50 to 150 pm shorter in the presence of water. Increasing the Cs$^{+}$ coverage from 0.16 to 0.48~ML at a given water coverage reduces both, the work function and the Cs$^{+} 6s$ energy, as shown in panels (b,c). A computed phase diagram, see Fig.\ S10, shows the most stable clusters. It demonstrates that small changes in concentration affect the cluster size suggesting that the cluster structure changes depending on the local concentration of Cs$^{+}$ and water, which complements the changes in the electronic structure discussed above.

\begin{figure*}[t]
\includegraphics[width=0.99\textwidth]{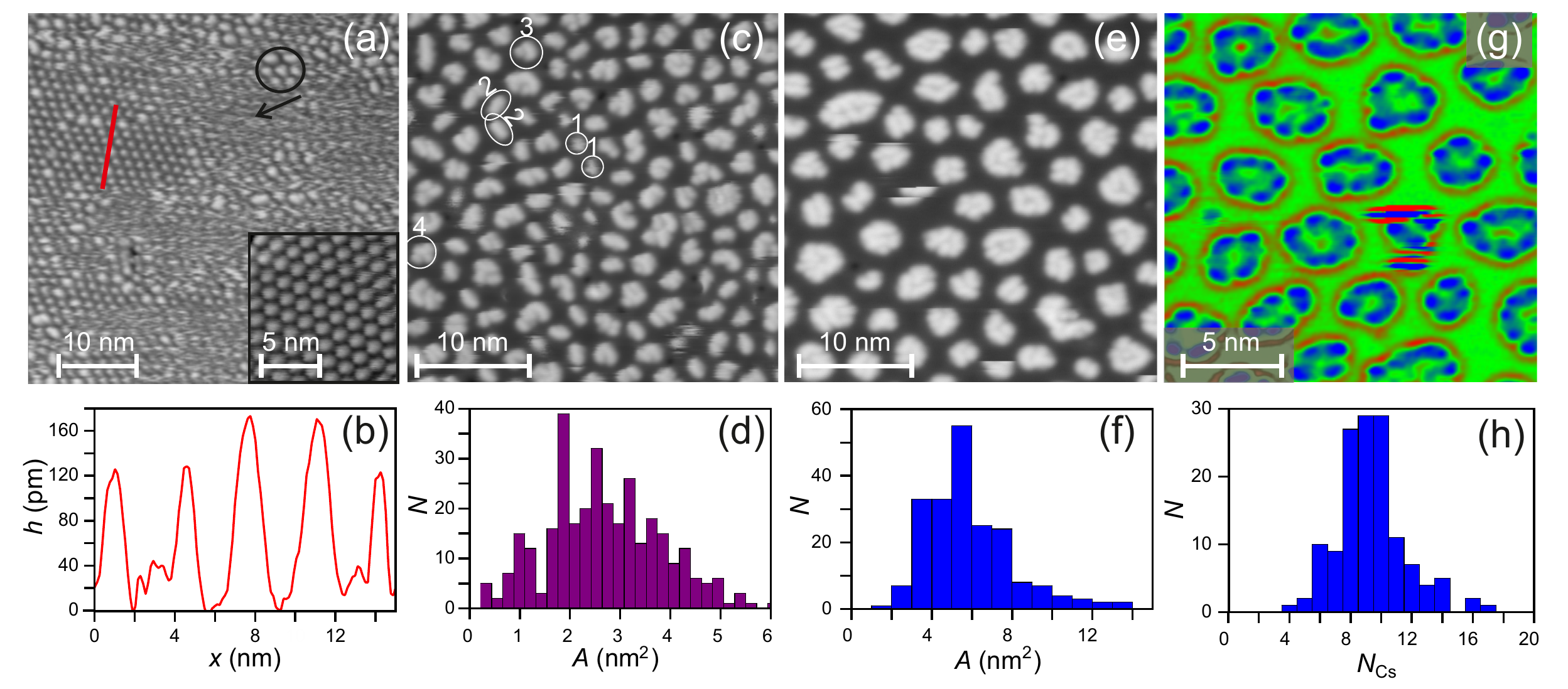}
\caption{Scanning tunnelling microscopy of Cs$^{+}$ ions coadsorbed on Cu(111) with D$_2$O: (a) initial hydration; circle surrounds some of the larger clusters; arrow points to a diffusive region; inset: Cs$^{+}$ only; (b) height profile along the line in (a); (c,d) STM image and area histogram of hydration at $\sim 10$ D$_2$O molecules per Cs$^{+}$ ion, with some clusters marked with the number of their subunits; (e-h) 10~D$_2$O molecules per Cs$^{+}$ ion annealed at 50~K for 15 min, (e,g) STM images on grey and Laplace-filtered on false-color scale ({\it cf}.\ Fig.~S1), (f) area histogram, (h) histogram of estimated number $N_{\mathrm{Cs}}$ per cluster; tunneling parameters:(a) -250~mV, 10~pA, (c) 53~mV, 7.5~pA, (e) 40~mV, 7.5~pA, (g) 100~mV, 10 pA.
\label{image_stm}}
\end{figure*}

To verify the calculated cluster structures, we used STM, see Supplemental Material for details. Starting with Cs deposited without water, a hexagonal superstructure forms, see the inset of Fig.\ \ref{image_stm}(a), at a coverage of 0.2~ML Cs$^{+}$, in agreement with \cite{von_hofe_adsorption_2006}. With a distance of $(1.55\pm0.07)$ nm between the protrusions this corresponds to a (6$\times$6) superstructure with respect to the hexagonal Cu(111) surface  suggesting long-range interaction between individual Cs$^{+}$, which was attributed to electrostatic repulsion \cite{von_hofe_adsorption_2006}. The striped appearance of some ions within the Cs$^{+}$ layer reflects their mobility even at $T=7$~K, {\it cf}.\ Fig.~S2. For an incompletely hydrated layer, parts of the surface are still covered by the hexagonal layer, upper left in Fig.\ \ref{image_stm}(a). Here, protrusions in the hexagonal array are imaged broader, at an area of $(0.34 \pm 0.08)$ nm$^2$ as compared to  $(0.83 \pm 0.09)$ nm$^2$, and higher, at 170~pm instead of 125~pm, Fig.\ \ref{image_stm}(b). Their distinct size suggests that only one water molecule is attached to a single ion. On other parts of the surface wider clusters exist, see circle in Fig.\ \ref{image_stm}(a). Here, the distance between the clusters is larger than in the hexagonal array. These regions are surrounded by some diffusive layer, indicative of mobile Cs$^{+}$. Such a mobility is possible if there is more than one Cs$^{+}$ ion bound in each cluster, and each ion occupies less space than before solvation. For such an approach of Cs$^{+}$ ions, the water needs to screen the ions and compensate their repulsive Coulomb interaction by bonding.

At higher water coverages of $\sim 10$ water molecules per Cs$^{+}$, all Cs$^{+}$ ions are bound within D$_2$O-Cs$^{+}$ clusters, see Fig.\ \ref{image_stm}(c). The considerably decreased number of D$_2$O-Cs$^{+}$ clusters as compared to the original number of Cs$^{+}$ ions, confirms that more than one Cs$^{+}$ is bound within each cluster; on average we identify four to five. Elongated clusters are frequent and are marked by ellipsoids '2'. These lead to a distinct maximum at an area of $(1.9 \pm 0.2)$ nm$^2$ in the area histogram in Fig.\ \ref{image_stm}(d), which suggests clusters of around half and 1.5 times this size. Their apparent size in the STM image as compared to sizes of pure water clusters is consistent with 10~water molecules per Cs$^{+}$ ion \cite{michaelides_ice_2007}, suggesting that these smallest units contain one Cs$^{+}$ ion. The smallest clusters thereby show the characteristic stripes of a mobile species, marked by circles '1' in panel (c). This is in contrast to immobile, larger clusters which consist of subunits of this size, leading to distinct multiples in the area histogram. Thereby, clusters with three protrusions are triangular, those with four protrusions rectangular. Larger clusters are far from being uniform in shape.

Similar structures were found after increasing the temperature. Upon annealing at 50~K, the cluster size increases, Fig.\ \ref{image_stm}(e). The mean area doubles from $(2.7 \pm 1.2)$~nm$^2$ to $(5.9 \pm 2.3$)~nm$^2$, see Fig.\ \ref{image_stm}(f). The cluster size is far from uniform, but all clusters seem to consist of subunits that align along their perimeter. These subunits are enhanced in visibility by color-coding a Laplace-filtered image in Fig.\ \ref{image_stm}(g), see Fig.~S1 for a comparison of regular and filtered images. Assuming that each of the circular blue dots of smallest height contains one Cs$^{+}$ ion and that the elongated or the higher protrusions contain two Cs$^{+}$ ions leads to a number of Cs$^{+}$ ions per area equivalent to the deposited value. The corresponding histogram reveals that 85\% of the clusters contain 8 to 10 Cs$^{+}$ ions, a rather uniform distribution. Thus, the STM results at high water coverage confirm the structure found by DFT. Instead of a central ion with a solvation shell, we observe water clusters with multiple ions at the perimeter, an arrangement which we term "inside-out hydration".

\begin{figure}[t]
\includegraphics[width=0.95\columnwidth]{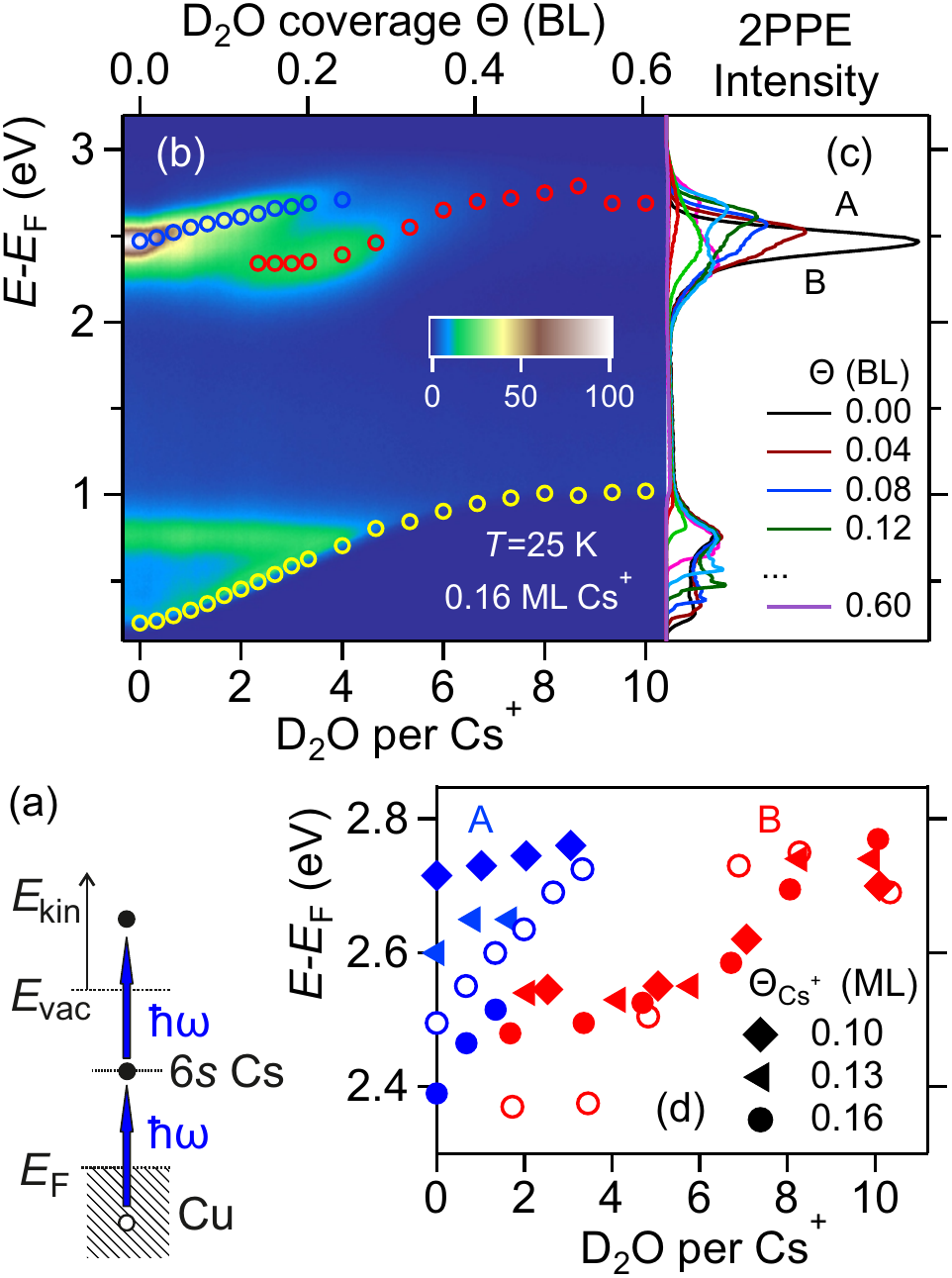}
\caption{
(a) Schematic of 2PPE on Cs$^{+}$/Cu(111). (b) False color map of 2PPE intensity as a function of D$_2$O coverage and $E-E_{\mathrm{F}}$. The D$_2$O coverage is given in bilayers (top axis) and as number of molecules D$_2$O per Cs$^{+}$ (bottom axis), see Supplemental Material for details. The Cs $6s$ states dressed with D$_2$O until 2.7 D$_2$O/Cs$^{+}$, for $\Theta > 0.16$~BL or 2.7 D$_2$O/Cs$^{+}$, and the change in work function are depicted by blue, red, and yellow circles, respectively. (c) 2PPE spectra at selected $\Theta$. (d) The change in energy of the two states (at low water coverage, {\textbf A}, in blue, and at high water coverage, {\textbf B}, in red) as a function of D$_2$O/Cs$^{+}$ for different Cs$^{+}$ coverage $\Theta _{{\mathrm Cs}^{+}}$ and $T=25$~K and 80 K depicted by open and closed symbols, respectively.
\label{image_2ppe}}
\end{figure}

Having established a good agreement between calculated and experimental cluster structures, we turn now to the electronic structure. Experimentally, this was studied by 2PPE, for which we employ femtosecond laser pulses of photon energy $\hbar \omega$=3.1~eV \cite{sandhofer_unoccupied_2014,thomas_competition_2021}. In 2PPE, the first photon $\hbar \omega$ excites resonant electron transfer from Cu(111) to the unoccupied Cs $6s$ state. A second photon generates the photoelectron analyzed in a spectrometer \cite{sandhofer_unoccupied_2014}, see Fig.\ \ref{image_2ppe}(a). To study the interaction of D$_2$O with Cs$^{+}$, we acquired 2PPE spectra while adsorbing D$_2$O on Cs$^{+}$/Cu(111) at 25~K as shown in Fig.\ \ref{image_2ppe}(b,c). The peaks at energies $E-E_{\mathrm{F}}$=2.47~eV and 0.75~eV observed for bare Cs$^{+}$/Cu(111) are assigned to the unoccupied Cs $6s$ state \cite{zhao_electronic_2008} and the occupied $3d$-band of the Cu(111) substrate, respectively. Here, $E_{\mathrm{F}}$ is the Fermi energy of Cu(111). The energy of the Cs $6s$ state increases during the adsorption of D$_2$O by 240~meV, see Fig.\ \ref{image_2ppe}(b). The intensity of the Cs $6s$ state decreases with D$_2$O coverage and vanishes at 4~D$_2$O molecules per ion. Slightly below this coverage a new state appears at 2.34~eV. Its energy increases with D$_2$O coverage as well, see red circles. This state may be solvent-dependent, as it was not observed for Cs$^{+}$-Xe on Cu(111) \cite{thomas_competition_2021}. The change in the work function of the surface is manifested as a change in the low energy cut-off of spectra, see yellow circles.

Fig.\ \ref{image_2ppe}(d) compiles energies for different Cs$^{+}$ coverages $\Theta _{{\mathrm Cs}^{+}}$ and at two water adsorption temperatures, $T$=25~K and 80~K. In agreement with Ref.~\cite{zhao_electronic_2008}, the energy of the Cs $6s$ state for bare Cs$^{+}$/Cu(111) decreases with increasing $\Theta _{{\mathrm Cs}^{+}}$. The unoccupied Cs $6s$ state dressed with D$_2$O is designated as {\textbf A} (blue markers). The unoccupied Cs $6s$ derived state observed for more water is denoted as {\textbf B} (red markers). Upon adsorption of D$_2$O, the energy of the {\textbf A} state increases. Remarkably, the energy of the {\textbf B} state is independent of the Cs$^{+}$ coverage but increases with D$_2$O coverage by more than 200~meV.  At  $T$=25~K, the energy of the {\textbf B} state is nearly constant below 4~D$_2$O/Cs$^{+}$ and above 7~D$_2$O/Cs$^{+}$, and the increase in energy occurs via a jump at around 6 to 7  D$_2$O/Cs$^{+}$. This is in contrast to the weaker, more gradual increase at $T$=80~K, which indicates limited mobility of Cs-water clusters at the lower temperature.

These results agree qualitatively with our DFT calculations, but the latter overestimate the observed shifts by a factor of 2, {\it cf}.\ Figs.\ \ref{image_dft},\ref{image_2ppe}. The appearance of two different resonance states at water coverages around 1~D$_2$O/Cs$^{+}$ suggests that different clusters coexist, in agreement with the STM results. To test this hypothesis, we performed calculations using a larger 7$\times$7 unit cell. This enables us to investigate the electronic structures of different clusters within one unit cell. As shown in Fig.\ S6, the maxima of the unoccupied Cs$^{+} 6s$ bands of Cs$^{+}$ ions in different clusters have different energies.

This agreement among experiment and theory allows us to propose a scheme for the hydration of Cs$^{+}$ on Cu(111). At low water coverage ($\leq$ 1 water per Cs$^{+}$), Cs$^{+}$ ions are dispersed on the surface in a hexagonal array and are bonded to at most one water molecule. Adding water leads to the formation of small clusters with low water:Cs$^{+}$ ratio (up to ca.\ 3:1), which shifts the $6s$ state to higher energies. At a coverage of three water molecules per Cs$^{+}$ ion, larger clusters with multiple Cs$^{+}$ ions start to dominate. They show a new lower-energy $6s$ state due to the higher local Cs$^{+}$ coverage. Adding water shifts this state to higher energies.

The hydration structure of Cs$^{+}$ on Cu(111) is markedly different compared to bulk water or the gas phase. Due to adsorbate-surface interactions, two-dimensional clusters are energetically preferred. Together with the large size of Cs$^{+}$ ions, this leads to a competition between Cs$^{+}$-water bonds and hydrogen bonds. The latter dominate in the most stable cluster structures, leading to a preference for water clusters with Cs$^{+}$ ions at their perimeter.

As demonstrated here for a highly crowded situation, the relationship between coverage, cluster structure, and unoccupied electronic states may have important consequences for the reactivity of solvated alkali/metal interfaces in heterogeneous catalysis and related fields. Such understanding of the elementary interactions that determine the structures of ion-solvent clusters may also provide useful input for an advanced model description of electrode-electrolyte interfaces. While the peculiar ``inside out'' solvation structure may not be visible at high water coverages, the balance of the fundamental interactions shown could still play an important role in the chemistry and physics of solvated alkali ions at the metal-liquid interface.

\section*{Acknowledgments}
Funded by the Deutsche Forschungsgemeinschaft (DFG, German Research Foundation) under Germany's Excellence Strategy - EXC 2033 - 390677874 - RESOLV (C.B., U.B, K.M., J.T.) as well as EXC 2008/1-390540038 - UniSysCat (C.P. and P.S.). The DFG is furthermore acknowledged for funding within Project ID No. 278162697-SFB 1242 (J.T.). C.P. is grateful to the Alexander von Humboldt foundation for financial support within the Feodor Lynen program. We thank M. Meyer for experimental support, as well as M. Wolf and A. Rubio for fruitful discussions.

\end{document}